**An energy harvesting technology controlled by ferromagnetic resonance**

Yuta Nogi,[1] Yoshio Teki,[2] and Eiji Shikoh[1,a)]

[1] *Graduate School of Engineering, Osaka City University, 3-3-138 Sugimoto, Sumiyoshi-ku, Osaka 558-8585, Japan*

[2] *Graduate School of Science, Osaka City University, 3-3-138 Sugimoto, Sumiyoshi-ku, Osaka 558-8585, Japan*

We have successfully demonstrated electrical charging using the electromotive force (EMF) generated in a ferromagnetic metal (FM) film under ferromagnetic resonance (FMR). In the case of $Ni_{80}Fe_{20}$ films, electrical charge due to the EMF generated under FMR can be accumulated in a capacitor; however, the amount of charge is saturated well below the charging limit of the capacitor. Meanwhile in the case of $Co_{50}Fe_{50}$, electrical charge generated under FMR can be accumulated in a capacitor and the amount of charge increases linearly with the FMR duration time. The difference between the $Ni_{80}Fe_{20}$ and $Co_{50}Fe_{50}$ films is due to the respective magnetic field ranges for the FMR excitation. When the FM films were in equivalent thermal states during FMR experiments, $Co_{50}Fe_{50}$ films could maintain FMR in a detuned condition, while $Ni_{80}Fe_{20}$ films were outside the FMR excitation range.

The EMF generation phenomenon in an FM film under FMR can be used an energy harvesting technology by appropriately controlling the thermal conditions of the FM film.

a) E-mail: shikoh@eng.osaka-cu.ac.jp

Energy harvesting is an important technology to efficiently utilize the earth's natural resources.[1] This technology harvests the existing micro-energy in an environment, and is different from conventional electric generation technologies such as electric power plants. So far the harvesting of such micro-energies has focused on the use of light, heat, vibration, electromagnetic fields, and their related phenomena.[1] The energy obtained per system due to such harvesting methods is not very large; however, the harvested electric power has the potential to be used to operate electronic devices.

Ferromagnetic resonance (FMR) is a magnetic phenomenon in which the magnetization dynamics in a magnetic material is controlled using both a static magnetic field ($H$) and a high frequency magnetic field in the GHz band.[2] In research on spintronics, it has been discovered that an electromotive force (EMF) is generated in the ferromagnetic metal (FM) film itself under FMR.[3,4] The EMF originates from various physical phenomena such as the inverse Hall effect (ISHE),[3-5] the anomalous Hall effect (AHE),[3-5] and so on. In conventional devices, the EMF generated in the FM film under FMR must be carefully removed, for example, in spin injection and spin transport experiments by the spin-pumping driven by the FMR.[5-14] Meanwhile, the EMF generation phenomenon itself in an FM film under FMR is the focus of this study, independent of the EMF origins. We have conceived an energy harvesting technology which uses this EMF generation phenomenon under FMR. We have successfully demonstrated electrical charging using the EMF generated in an FM film itself under the FMR, and show that the EMF generation phenomenon under the FMR is usable as an energy harvesting technology with appropriate control of the thermal conditions of the FM film.

Figure 1(a) shows a schematic illustration of our sample structure and the experimental set-up to detect the EMF generated in the sample under FMR. An FM film was formed on a thermally-oxidized silicon substrate using an electron beam deposition system at pressure <$10^{-6}$ Pa. $Ni_{80}Fe_{20}$ (Kojundo Chemical Lab. Co., Ltd., 99.99% purity) or $Co_{50}Fe_{50}$ (Kojundo Chemical Lab. Co., Ltd., 99.99%) was used as the FM film. The deposition rate and the substrate temperature during FM deposition were set to 0.03 nm/s and room temperature (RT), respectively. No cover layer to prevent the FM films from oxidizing was formed because only the EMF phenomenon generated in the FM films under the FMR was considered in this study, not the individual origins of EMFs. After the FM deposition, the sample substrates were cut to the designed size as shown in Fig. 1(a).

To confirm the EMF properties generated in the FM films under FMR, a microwave $TE_{011}$-mode cavity in an electron spin resonance system (JEOL, JES-TE300) was used to excite the FMR in an FM film, and a nanovoltmeter (Keithley Instruments, 2182A) was used to measure the EMF. Lead wires to detect the output voltage properties were directly attached at both ends of the FM film with silver paste.

To evaluate the electric charging properties under the FMR, the electrical circuit shown in Fig. 1(b) was connected to the FM film sample, in place of the nanovoltmeter used for the above EMF confirmation. First, all of the switches $S_1$, $S_2$ and $S_3$ were opened, and the capacitor (the capacitance is $C$) was completely discharged. In the case of charging experiments, $S_1$ and $S_2$ are closed, and the FMR of the FM film was excited by the same ESR system as described above. The electrical current derived from the EMF generated in the FM film under the FMR flowed and the electric charges were accumulated in the capacitor for the FMR duration time. The FMR condition in an in-plane field was set according to Kittel's formula[15]

$$\frac{\omega}{\gamma} = \sqrt{H_{FMR}(H_{FMR} + 4\pi M_S)}, \qquad (1)$$

where $\omega$, $\gamma$, $H_{FMR}$ and $M_S$ are the angular frequency ($2\pi f$), the gyromagnetic ratios of $1.86 \times 10^7$ G$^{-1}$s$^{-1}$ for $Ni_{80}Fe_{20}$,[2,8] and $1.84 \times 10^7$ G$^{-1}$s$^{-1}$ for $Co_{50}Fe_{50}$ calculated from the g-factor,[16] the FMR field and the saturation magnetization of the FM film, respectively. After the charging processes, both $S_1$ and $S_2$ were opened.

Next, the accumulated charges in the capacitor were discharged using a so-called RC-series circuit, where the accumulated charges in the capacitor are consumed in a resistor (the resistance is $R$) as heat. Before starting discharge experiments for evaluation of the amount of charge accumulated in the charging processes, $S_3$ was closed, that is, the same nanovoltmeter as above was connected. The trigger for discharge is $S_2$. When $S_2$ is closed, an electric current due to the charge accumulated in the $C$ starts to flow and is consumed at the resistor. In the RC-series circuit shown in Fig. 1(b), the electrical voltage between the terminals of the resistor is defined as $V(t)$, which is described by the

following equation:

$$V(t) = V_0 exp\left(-\frac{t}{\tau}\right), \qquad (2)$$

where $V_0$, $t$, and $\tau$ are the initial voltage corresponding to the accumulated charge during the FMR duration time, the duration time from the trigger of the discharge process, and the time constant of the discharge circuit, which is defined as $RC$. In this study, $\tau$ was always set to be 1 s for ease of measurement (for example, $R : C = 1\ \Omega : 1$ F). All evaluations were performed at RT.

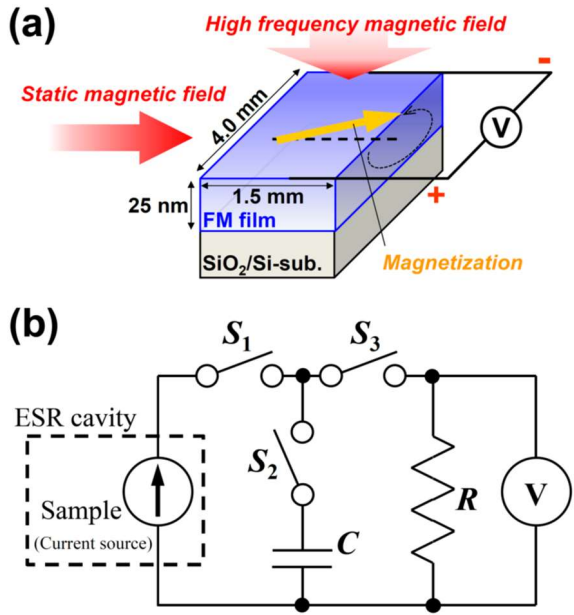

FIG. 1. (a) A schematic illustration of our sample structure and the experimental set-up to detect the EMF generated in the sample under FMR., (b) Electrical circuit to evaluate the electric charging properties.

Figure 2(a) and (b) show the FMR spectrum of an $Ni_{80}Fe_{20}$ film and the EMF generated in the same $Ni_{80}Fe_{20}$ film itself under the FMR at the microwave frequency and power ($P$) of 9.45 GHz and 200 mW, respectively. In Fig. 2(b), the circles represent experimental data and the solid line is the fitted curve obtained using the following equation[3-5,8-14]

$$V(H) = V_{Sym}\frac{\Gamma^2}{(H-H_{FMR})^2+\Gamma^2} + V_{Asym}\frac{-2\Gamma(H-H_{FMR})}{(H-H_{FMR})^2+\Gamma^2}, \quad (3)$$

where $\Gamma$ denotes the damping constant (26 Oe for $Ni_{80}Fe_{20}$ in this study). The first and second terms in Eq. (3) correspond to the symmetry term for $H$ due to the ISHE, and the asymmetry term for $H$ due to the AHE and/or other effects showing the same asymmetric voltage behavior relative to $H$, respectively.[3-5,8-14] $V_{Sym}$ and $V_{Asym}$ correspond to the coefficients of the first and second terms in the Eq. (3). The $H_{FMR}$ of the $Ni_{80}Fe_{20}$ film was 1,100 Oe and the $M_S$ of the $Ni_{80}Fe_{20}$ film was estimated to be 646 emu/cc with Eq. (1). The output voltages from the $Ni_{80}Fe_{20}$ film under the FMR are observed at $H_{FMR}$. The observed EMF is mainly due to the self-induced ISHE in the $Ni_{80}Fe_{20}$ film under FMR.[3]

Figure 2(c) and (d) show the FMR spectrum of a $Co_{50}Fe_{50}$ film and the EMF generated in the same $Co_{50}Fe_{50}$ film itself under the FMR at the microwave frequency and $P$ of 9.45 GHz and 200 mW, respectively. In Fig. 2(d), the circles represent experimental data and the solid line is the fitted curve obtained using Eq. (3) with the $\Gamma$ of 110 Oe for $Co_{50}Fe_{50}$ in this study. The $H_{FMR}$ of the $Co_{50}Fe_{50}$ film was 572 Oe and the $M_S$ of the $Co_{50}Fe_{50}$ film was estimated to be 1410 emu/cc with Eq. (1). The FMR spectrum of the $Co_{50}Fe_{50}$ film is wider than that of the $Ni_{80}Fe_{20}$ film, corresponding to the difference in magnetic anisotropy. Output voltages from the $Co_{50}Fe_{50}$ film under the FMR are observed at $H_{FMR}$.

The origins of the EMF generated in a $Co_{50}Fe_{50}$ film under FMR are currently under investigation,[17] while these might come from the ISHE, the AHE and so on, similarly to other FM films, like $Ni_{80}Fe_{20}$, Fe and Co films.[3,4] While the apparent EMF generated in the $Co_{50}Fe_{50}$ film at the $H_{FMR}$ is smaller than that of the $Ni_{80}Fe_{20}$ film, it is sufficient for the energy-harvesting experiments in this study.

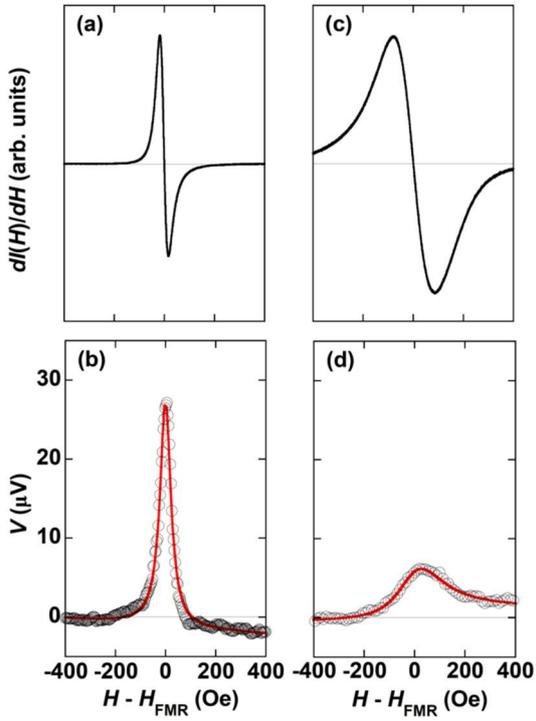

FIG. 2. (a) FMR spectrum of a $Ni_{80}Fe_{20}$ film and (b) the EMF generated in the same $Ni_{80}Fe_{20}$ film under FMR. (c) FMR spectrum of a $Co_{50}Fe_{50}$ film and (d) the EMF generated in the same $Co_{50}Fe_{50}$ film under FMR. The microwave frequency and power are 9.4 GHz and 200 mW, respectively.

The energy-harvesting experiments are described below. First, using the charging circuit, the electric current generated in the FM films under the respective FMR condition to satisfy Eq. (1) flows, and the capacitor is charged. The microwave power was 200 mW in all experiments except for the evaluation of $P$-dependence. Each FMR excitation was maintained for 30 min. with the FMR condition to satisfy the Eq. (1), and then, the capacitor was discharged. Figure 3(a) shows typical discharge properties of the capacitor evaluated using the discharge circuit. Circles and triangles are experimental data for a $Ni_{80}Fe_{20}$ film and a $Co_{50}Fe_{50}$ film, respectively. The solid lines are fitted curves obtained using Eq. (2), and the respective data showed a good fit. The $V_0$ is about 78 µV for the $Ni_{80}Fe_{20}$ film and 56 µV for the $Co_{50}Fe_{50}$ film. Figure 3(b) shows the $P$ dependences of the discharge properties for a $Ni_{80}Fe_{20}$ film. Each FMR duration time (the charging time) was 30 min. The solid lines are fitted curves obtained using Eq. (2). Figure 3(c) shows the $P$ dependence of the $V_0$ analyzed from the Fig. 3(b). The value of $V_0$ increases linearly with the increase in $P$, that is, the charging is clearly due to the FMR phenomena.

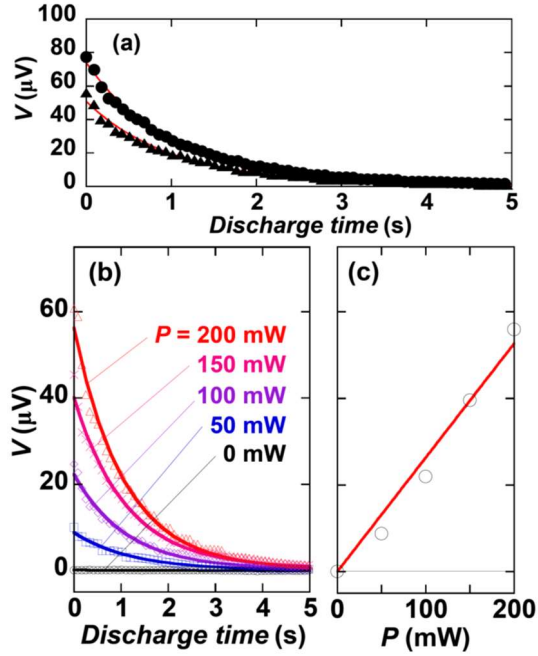

FIG. 3. (a) Typical discharge properties of capacitors evaluated using the discharge circuit. Circles and triangles are experimental data for a $Ni_{80}Fe_{20}$ film and a $Co_{50}Fe_{50}$ film, respectively. (b) Microwave power ($P$) dependences of discharge properties for a $Ni_{80}Fe_{20}$ film. Each FMR duration time (the charging time) was 30 min. The solid lines are fitted curves obtained using Eq. (2). (c) The $P$ dependence of the $V_0$ obtained by analysis of FIG. 3 (b).

Figure 4 shows the FMR duration time dependence of the discharge properties of a $Ni_{80}Fe_{20}$ film. The solid lines are fitted curves obtained using Eq. (2). Figure 4(b) shows the FMR duration time dependence of the $V_0$ generated in the $Ni_{80}Fe_{20}$ film, from analysis of Fig. 4(a). The amount of charge from the $Ni_{80}Fe_{20}$ film tends to be saturated well below the voltage limit of the capacitor when the FMR duration time is over 15 min. To investigate the reason why the charge is saturated, we changed the capacitors and resistors while keeping $\tau$ at 1s. Also, other $Ni_{80}Fe_{20}$ films were tested. However, in all harvesting experiments with $Ni_{80}Fe_{20}$ films, the charge was saturated against the FMR duration time. Therefore, we changed the FM film from $Ni_{80}Fe_{20}$ to another FM.

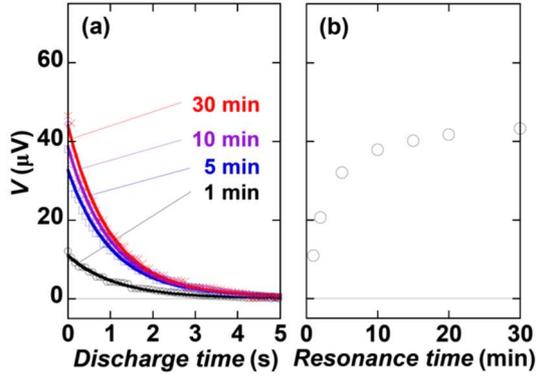

FIG. 4. (a) FMR duration time dependence of the discharge properties of a $Ni_{80}Fe_{20}$ film. The solid lines are fitted curves obtained using Eq. (2). (b) The FMR duration time dependence of the $V_0$ generated in the $Ni_{80}Fe_{20}$ film from analysis of FIG. 4(a).

Figure 5 shows the FMR duration time dependence of the discharge properties of a $Co_{50}Fe_{50}$ film. The solid lines are fitted curves obtained using Eq. (2). Figure 5(b) shows the FMR duration time dependence of the $V_0$ generated in the $Co_{50}Fe_{50}$ film obtained by analysis of Fig. 5(a). The amount of charge from the $Co_{50}Fe_{50}$ film increases almost linearly with increasing FMR duration time and is not saturated. This behavior is different from the case of the $Ni_{80}Fe_{20}$ film shown in the Fig. 4(b), and this characteristic shows good reproducibility with other $Co_{50}Fe_{50}$ films. It is noticed that while the apparent EMF generated in the $Co_{50}Fe_{50}$ film at the $H_{FMR}$ is smaller than that in the $Ni_{80}Fe_{20}$ film, and the FMR spectrum of a $Co_{50}Fe_{50}$ film is much wider than that of a $Ni_{80}Fe_{20}$ film. The FM film under FMR is basically heated. Notably, the FMR duration time in this study is very long compared with general FMR experiments. The $M_S$ of an FM film becomes smaller at high temperature than at low temperature. That is, as shown in the Eq. (1), the $H_{FMR}$ of an FM film becomes larger at high temperature than at low temperature when the microwave frequency is the same. Thus, in the experiments with $Ni_{80}Fe_{20}$ films, the film was heated and the $H_{FMR}$ shifted to larger values compared with the beginning of the experiment. Because the ESR-system parameters were kept the same in the experiments, the $Ni_{80}Fe_{20}$ film might have gone out of the FMR excitation range. Therefore, EMF was hardly generated and the charging of the capacitor almost stopped. Meanwhile, in the experiments with $Co_{50}Fe_{50}$ films, the film was heated under the FMR and the $H_{FMR}$ also became larger compared with the beginning of the experiment. Similarly, the ESR-system settings were maintained in the experiments. However, the $Co_{50}Fe_{50}$ film might not be fully out of the FMR excitation range for $Co_{50}Fe_{50}$ films, due to the wider FMR spectrum than $Ni_{80}Fe_{20}$. Therefore, the EMF generation in $Co_{50}Fe_{50}$ films under FMR was kept in a detuned state and the charging to the capacitor was maintained. Of course, thermally equivalent conditions for similar experiments must depend on the microwave cavity used and other thermal factors. That is, by appropriately controlling the thermal conditions around the FM film, the charging was maintained. The above result indicates that energy harvesting experiments using FMR were successfully demonstrated.

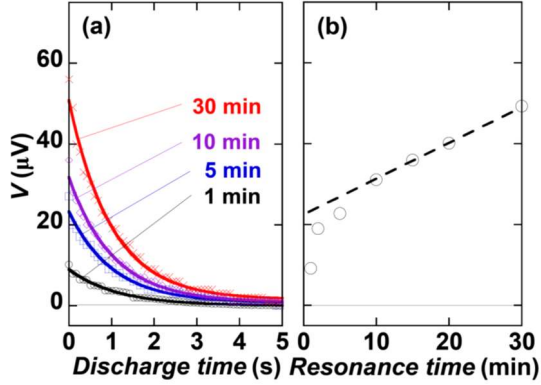

FIG. 5. The FMR duration time dependence of discharge properties for a $Co_{50}Fe_{50}$ film. The solid lines are fitted curves obtained using Eq. (2). (b) The FMR duration time dependence of $V_0$ generated in the $Co_{50}Fe_{50}$ film from analysis of the data in FIG. 5(a). The dashed line is a guide for the eyes.

At present, no diodes were connected to the circuit to rectify the electrical currents. To efficiently charge the capacitor and to increase the amount of accumulated charge, the use of diodes may be effective because the currents generated by an FM film under the FMR are very small, and controlling the flow of such micro-currents is usually difficult. The microwave power was kept the same (200 mW) in all experiments except for the evaluation of $P$-dependence. A smaller microwave power may be preferable to reduce heating of the films and to precisely control the thermal conditions of the FM film under FMR. For FMR excitations, in general, a large electric power is required to apply the electric current to provide a static magnetic field and a high frequency field. Thus, methods should be developed to reduce the electric power required for the excitation of FMR, using, for example, permanent magnets to create the static magnetic field and environmental electromagnetic waves for a high frequency magnetic field. While those might be hard to be establish, such technology is eagerly awaited because a lot of GHz-band microwaves exist in modern environments such as those used in wireless internet services. The above issues must be solved for practical use.

In summary, we successfully demonstrated electrical charging using the EMF generated in a FM film under FMR. In the case of $Ni_{80}Fe_{20}$, electrical charge due to the EMF generated under the FMR was stored in a capacitor; however, the amount of charge was saturated well below the charging limit of the capacitor despite the increase in FMR duration time. Meanwhile in the case of $Co_{50}Fe_{50}$, electrical charge generated under the FMR was stored in a capacitor and the amount of charge increased linearly with the FMR duration time. The FMR spectrum of the $Co_{50}Fe_{50}$ films was wider than that of $Ni_{80}Fe_{20}$. In equivalent thermal states during FMR experiments, $Co_{50}Fe_{50}$ films maintained FMR in a detuned condition, while $Ni_{80}Fe_{20}$ films were outside of the FMR excitation range. The above result indicated that the EMF generation phenomenon in FM films under FMR might be usable as an energy harvesting technology, by appropriately controlling the thermal conditions of the FM films.

This research was partly supported by a research grant from The Murata Science Foundation and a research grant from The Mazda Foundation.